\documentclass[aps,prl,twocolumn,showpacs,preprintnumbers,amsmath,amssymb,superscriptaddress]{revtex4}%

\usepackage{graphicx}%
\usepackage{dcolumn}
\usepackage{amsmath}

\makeatletter
\def\btt#1{\texttt{\@backslashchar#1}}%
\DeclareRobustCommand\bblash{\btt{\@backslashchar}}%
\makeatother

\begin{document}


\title{Pressure induced softening of YB$_6$:
pressure effect on the Ginzburg-Landau parameter
$\kappa=\lambda/\xi$ }

\author{R.~Khasanov}
\affiliation{Physik-Institut der Universit\"{a}t Z\"{u}rich,
Winterthurerstrasse 190, CH-8057, Z\"urich, Switzerland}
\author{P.S.~H\"afliger}
\affiliation{Physik-Institut der Universit\"{a}t Z\"{u}rich,
Winterthurerstrasse 190, CH-8057, Z\"urich, Switzerland}
\author{N.~Shitsevalova}
\affiliation{Institute for Problems of Materials Science, National
Academy of Science of Ukraine, 03680 Kiev, Ukraine}
\author{A.~Dukhnenko}
\affiliation{Institute for Problems of Materials Science, National
Academy of Science of Ukraine, 03680 Kiev, Ukraine}
\author{H.~Keller}
\affiliation{Physik-Institut der Universit\"{a}t Z\"{u}rich,
Winterthurerstrasse 190, CH-8057, Z\"urich, Switzerland}

\begin{abstract}
Measurements of the transition temperature $T_c$, the second
critical filed $H_{c2}$ and the magnetic penetration depth $\lambda$
under hydrostatic pressure (up to 9.2~kbar) in the YB$_6$
superconductor were carried out. A pronounced and {\it negative}
pressure effects (PE) on $T_c$ and $H_{c2}$ with
$dT_c/dp=-0.0547(4)$~K/kbar and $\mu_0dH_{c2}(0)/dp
=-4.84(20)$~mT/kbar, and zero PE on $\lambda(0)$ were observed.  The
PE on the coherence length $d\xi(0)/dp=0.28(2)$~nm/kbar was
calculated from the measured pressure dependence of $H_{c2}(0)$.
Together with the zero PE on the magnetic penetration depth
$\lambda(0)$, our results imply that the Ginzburg-Landau parameter
$\kappa(0)=\xi(0)/\lambda(0)$ depends on pressure and that pressure
''softens`` YB$_6$, e.g. moves it to the type-I direction.

\end{abstract}
\pacs{74.70.Dd, 74.62.Fj, 74.25.Ha}

\maketitle

The Ginzburg-Landau parameter $\kappa=\lambda/\xi$ ($\lambda$ is the
magnetic penetration depth and $\xi$ is the coherence length) is one
of the fundamental parameters for superconductors. The parameter
$\kappa$ establishes the border between type-I ($\kappa<1/\sqrt{2}$)
and type-II ($\kappa>1/\sqrt{2}$) superconductors by determining
point where the surface energy, associated with the domain wall
between the superconducting and the normal state areas, changes
their sign from "plus" to "minus".
Remarkably, two physical quantities entering $\kappa$ ($\lambda$ and
$\xi$) depend on different properties of the superconducting
material. Indeed, for BCS superconductors the zero temperature
coherence length obeys the relation \cite{Tinkham75}
\begin{equation}
\xi(0)=\frac{\hbar\langle v_{\rm F} \rangle}{\pi \Delta(0)}.
 \label{eq:xi-BCS}
\end{equation}
Here, $\langle v_{\rm F} \rangle$ is the average Fermi velocity and
$\Delta(0)$ is the zero temperature value of the superconducting
energy gap. The zero-temperature penetration depth is given by
\cite{Chandrasekhar93,Hirschfeld93}
\begin{equation}
\frac{1}{\lambda^2(0)}=\frac{e^2}{\pi^2\hbar c^2} \oint_{S_F}ds|{\bf
v}_F(s)|,
 \label{eq:lambda}
\end{equation}
where the integral runs over the Fermi surface $S_F$. Assuming
spherical Fermi surface Eq.~(\ref{eq:lambda}) reduces to
\begin{equation}
\lambda^{-2}(0)\propto S_F \langle v_F\rangle.
 \label{eq:lambda_spherical}
\end{equation}
A quick glance to Eqs.~(\ref{eq:xi-BCS}) and
(\ref{eq:lambda_spherical}) reveal that $\lambda(0)$ is determined
by the {\it normal} state properties only [$S_F$ and $\langle
v_F\rangle$] while $\xi(0)$ is determined by both -- the {\it
normal} [$\langle v_F\rangle$] and the {\it superconducting}
[$\Delta(0)$] state properties of the material. This means that if
one would be able to affect only the superconducting (normal) state
properties, the Ginzburg-Landau parameter $\kappa$ will change. As a
consequence, superconductors can be driven to the more type-I or to
the more type-II directions.

Experiments under pressure open a possibility to probe this
phenomena.
The reason for this can be understood by considering the simple
metal superconductors, like Sn, In, Pb,  Al, {\it etc}., where the
conduction electrons possess $s$, or $p$ character. Since the $s$
and $p$ electrons in simple metals are nearly free, one expects
approximately $v_{F}\propto S_F \propto V^{+2/3}$ ($V$ is the unit
cell volume). For typical values of the bulk modulus
$B=400-800$~kbar \cite{webelements} it leads to $d\ln v_{F}/dp= d\ln
S_{F}/dp = 2/3\cdot B^{-1}=$0.08-0.17~\%/kbar. Note that this effect
is almost one order of magnitude smaller than the corresponding
pressure effect on $T_c$: Sn ($d\ln T_c/dp=-1.30$~\%/kbar), In
(--~1.12~\%/kbar),  Pb (--~0.51~\%/kbar), Al (--~2.47~\%/kbar)
\cite{Eiling81}. Taking into account this and the fact that within
BCS theory $\Delta(0)$ is simply proportional to $T_c$
($2\Delta(0)/k_{B}T_c=3.52$, see e.g. \cite{Tinkham75}),
Eq.~(\ref{eq:xi-BCS}) implies that the pressure shift of the
coherence length $\xi(0)$  is almost the same as the one of $T_c$
\begin{equation}
\frac{d\ln\xi(0)}{dp}=\frac{d\ln\langle v_{\rm F} \rangle}{dp} -
\frac{d\ln [\Delta(0)]}{dp} = \frac{2}{3}B^{-1} - \frac{d\ln
T_c}{dp} .
 \label{eq:delta-xi}
\end{equation}
From the other hand, the pressure effect on $\lambda(0)$ [see
Eq.~(\ref{eq:lambda_spherical})] is determined by the pressure
induced change of the Fermi surface and the Fermi velocity
\begin{equation}
\frac{d\ln\lambda(0)}{dp}=-\frac{1}{2}\frac{d\ln
S_F}{dp}-\frac{1}{2}\frac{d\ln \langle v_F\rangle}{dp} =
-\frac{2}{3}B^{-1},
 \label{eq:delta-lambda}
\end{equation}
which  appears to be an order of magnitude smaller as shown above.
Thus, the much bigger pressure effect (PE) on $\xi(0)$ in comparison
with the one on $\lambda(0)$ implies pressure dependence of the
Ginzburg-Landau coefficient $\kappa(0)$.

Here we report studies of the effect of pressure on the
Ginzburg-Landau coefficient $\kappa$ in YB$_6$ superconductor. It
was found that with increasing pressure from 0 to 9.2~kbar
$\kappa(0)$ decreases by an almost 8\% [from 6.17(5) to 5.70(4)],
implying that pressure drives YB$_6$ to the {\it type-I} direction.
The pressure effects on two quantities entering $\kappa(0)$
[$\lambda(0)$ and $\xi(0)$] were studied separately. It was obtained
that the PE on $\kappa(0)$ arises mostly from the pressure
dependence of the coherence length $\xi(0)$, while no PE on the
magnetic penetration depth $\lambda(0)$ (within the experimental
accuracy) was observed. It was also found that in BCS
superconductors for which the absolute value of the pressure shift
of $T_c$ is much bigger than $1/B$,  the pressure shifts of the
superconducting quantities such as $T_c$, $\kappa(0)$, $\xi(0)$,
$H_{c2}(0)$, $dH_{c2}/dT|_{T=T_c}$ are related to each other.


Details on the sample preparation for YB$_6$ single crystal can be
found elsewhere \cite{Lortz06}. The hydrostatic pressure was
generated in a copper--beryllium piston cylinder clamp especially
designed for magnetization measurements under pressure
\cite{Straessle02}. The sample was mixed with Fluorient FC77
(pressure transmitting medium). In both below described experiments
the sample-to-liquid volume ratio was of approximately $1/6$.

The field--cooled (FC) and zero-field cooled (ZFC) magnetization
measurements were performed with a SQUID magnetometer in fields
ranging from $ 0.5$~mT to 0.3~T and at temperatures between $1.75$~K
and $10$~K. Two sets of magnetization under pressure experiments
were performed. In the first one, small parts of the main single
crystal together with the small piece of In (pressure indicator)
were added to the pressure cell. In these experiments the pressure
dependences of the transition temperature $T_c$ and the upper
critical field $H_{c2}$ were studied. In the second set, the part of
the single crystal was grounded in mortar and then sieved via
10~$\mu$m sieve in order to obtain small grains needed for
determination of $\lambda$ from magnetization measurements.

\begin{figure}[htb]
\includegraphics[width=1.3\linewidth]{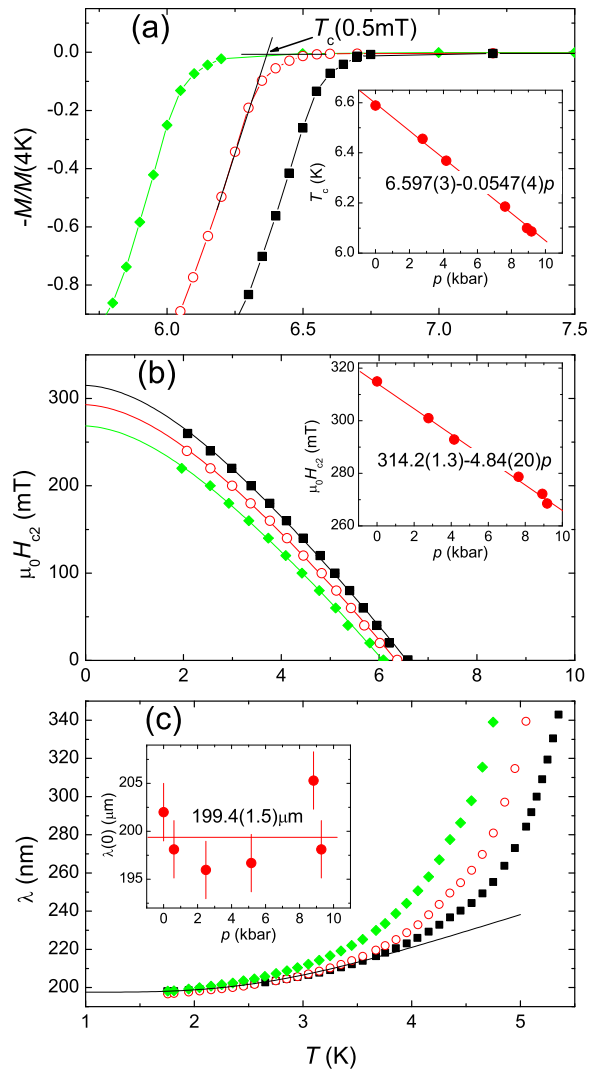}
\caption{(Color online) Temperature dependences of the low-field
(0.5~mT FC) magnetization (a), upper critical field $H_{c2}$ (b) and
magnetic penetration depth $\lambda$  measured at (from the right to
the left) 0.0, 4.16 and 9.18~kbar [panels (a) and (b)] and 0.60,
5.15 and 9.15~kbar [panel (c)]. The solid lines in (b) and (c)
correspond to the fit of the WHH (Ref.~\cite{Werthamer66}) and the
low-temperature s-wave BCS [Eq.~(\ref{eq:low_T_BCS})] models to the
experimental data, respectively. Inserts show the corresponding
pressure dependences of $T_c$ and the zero-temperature values of
$H_{c2}(0)$ and $\lambda(0)$ (see text for an explanation).}
 \label{fig:TcHc2Lam}
\end{figure}

Figure~\ref{fig:TcHc2Lam} shows the temperature dependences of the
low-field (0.5~mT FC) magnetization (a), upper critical field
$H_{c2}$ (b) and magnetic penetration depth $\lambda$ (c) measured
at different pressures. In the following we briefly discuss each
dependence separately.

With increasing pressure magnetization curves shifts almost parallel
to the lower temperatures implying the negative pressure effect on
the transition temperature $T_c$ [see Fig.~\ref{fig:TcHc2Lam}(a)].
The corresponding pressure dependence of $T_c$ is shown in the
inset. $T_c$ was taken from the linearly extrapolated $M(T)$ curves
in the vicinity of $T_c$ with the $M=0$ line. The linear fit yields
$dT_c/dp=-0.0547(4)$~K/kbar. Note that this value is in good
agreement with $dT_c/dp\simeq-0.053(8)$~K/kbar obtained indirectly
by Lortz {\it et al.} \cite{Lortz06} from thermal expansion
measurements.

The values of the second critical field $H_{c2}(T)$ presented in
Fig.~\ref{fig:TcHc2Lam}(b) were extracted from the FC magnetization
curves $M(T,H)$ measured in constant magnetic fields ranging from
0.5~mT to 0.3~T. For each particular field $H$ the corresponding
transition temperature $T_c(H)$ was determined as described above
[see Fig.~\ref{fig:TcHc2Lam}(a)]. The value of the field $H$ was
then attributed to be the upper critical field $H_{c2}$ at the
temperature $T=T_c$. The solid lines represent fits of the WHH model
\cite{Werthamer66} to the experimental data. The values of the upper
critical field at $T=0$ [$H_{c2}(0)$] obtained from the fits are
plotted in the inset as the function of pressure. The linear fit
yields $\mu_0dH_{c2}/dp=-4.84(20)$~mT/kbar. The values of $T_c$ and
$H_{c2}(0)$ measured at various pressures are summarized  in
Table~\ref{Table:pressure_results}.

The temperature dependences of $\lambda$ were calculated from the
measured 0.5~mT ZFC magnetization by using the Shoenberg formula
\cite{Shoenberg40}:
\begin{equation}
\chi  =  -\frac{3}{2}\left(1-
\frac{3\lambda}{R}\coth\frac{R}{\lambda}+
\frac{3\lambda^2}{R^2}\right),
 \label{eq:Shoenberg-modified}
\end{equation}
where $\chi=M/HV$ is the volume susceptibility, $V$ is the volume of
the sample and $R$ is the mean radius of the grains. The reducing of
the grain size with pressure was taken into account in $\lambda(T)$
calculation [Eq.~(\ref{eq:Shoenberg-modified})] by using the bulk
modulus value $B=1900$~kbar \cite{Zirngiebl86}. Due to unknown $R$,
the value of $\lambda$ at $T=1.7$~K and $p=0$ was normalized to
$\lambda(1.7$~K)=202~nm obtained from muon-spin rotation experiments
\cite{Khasanov06unp}. In order to obtain the zero temperature values
of $\lambda$ the low temperature part of the data were fitted by the
standard s-wave BCS model:
\begin{equation}
\frac{\Delta\lambda(T)}{\lambda(0)}=\sqrt{\frac{\pi\Delta(0)}{2k_BT}}
\exp\biggl(-\frac{\Delta(0)}{k_BT} \biggr).
 \label{eq:low_T_BCS}
\end{equation}
In order to decrease the number of the fitting parameters it was
assumed that $\Delta(0)$ scales with $T_c$ as
$2\Delta(0)=4.02k_BT_c$ \cite{Schneider87}. Note that the independence
of the ratio $2\Delta(0)/k_BT_c$ of pressure was recently confirmed
for RbOs$_2$O$_6$ \cite{Khasanov04} and ZrB$_{12}$ \cite{Khasanov05}
BCS superconductors. The resulting $\lambda(0)$ vs. $p$ dependence
is shown in the inset of the Fig.~\ref{fig:TcHc2Lam}(c). The linear
fit yields $d\lambda(0)/dp=0.22(33)$~nm/kbar. This implies that
within the accuracy of the experiment $\lambda(0)$ does not depend
on pressure. The mean value of $\lambda(0)$ was found to be
199.4(1.5)~nm.

\begin{figure}[htb]
\includegraphics[width=1.15\linewidth]{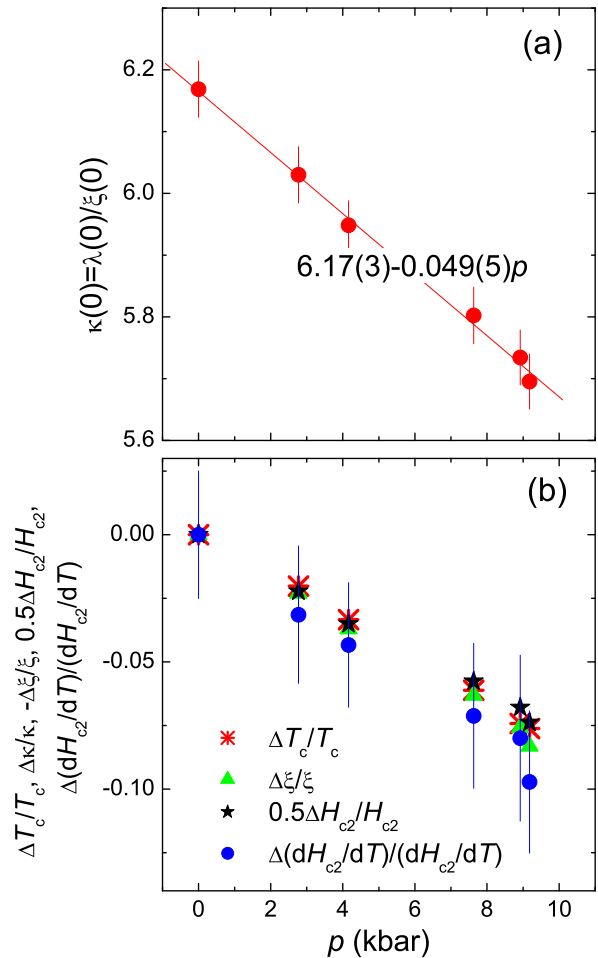}
\caption{(Color online) Pressure dependences of $\kappa(0)$ (a), and
the relative pressure shifts $\Delta T_c/T_c$,
$\Delta\kappa(0)/\kappa(0)$, $\Delta\xi(0)/\xi(0)$, $0.5 \Delta
H_{c2}(0)/H_{c2}(0)$ and
$\Delta(dH_{c2}/dT|_{T=T_c})/(dH_{c2}/dT|_{T=T_c})$ (b). The solid
line in (a) is a linear fit with the parameters shown in the Figure.
}
 \label{fig:relative_shifts}
\end{figure}

Using the obtained values of $H_{c2}(0)$ (see
Table~\ref{Table:pressure_results}), the Ginzburg-Landau coherence
length at T=0~K was calculated according to \cite{Tinkham75}:
\begin{equation}
\xi(0)=\sqrt{\frac{\Phi_0}{2 \pi H_{c2}(0)}},
 \label{eq:xi}
\end{equation}
where $\Phi_0$ is the magnetic flux quantum. The values of
$\kappa(0)$ were further calculated by using
$\lambda(0)=199.4(1.5)$~nm. It is seen (see
Fig.~\ref{fig:relative_shifts}) that $\kappa(0)$ decreases almost
linearly from 6.17(5) at ambient pressure to 5.70(4) at
$p=9.18$~kbar. The decrease of $\kappa$ thus implies that pressure
''softens`` the YB$_6$ superconductor by driving it to the type-I
direction.

\begin{table}[htb]
\caption[~]{\label{Table:pressure_results} Summary of the pressure
effect results (see text for details).
} %
\begin{center}
\begin{tabular}{ccccccccc}\\
 \hline
 \hline
  $p$&$T_{c0}$&$\mu_0 H_{c2}(0)$&$\mu_0dH_{c2}/dT$&$\xi(0)$ &$\kappa(0)$&\\
(kbar)&(K)&(mT)&(mT/K)&(nm)&&\\
\hline
0.0  &6.589(3) &315.0(1.7)&66.0(1.1) &32.32(17) &6.17(5)\\
2.77 &6.456(3) &301.0(1.7)&63.9(1.2) &32.07(19) &6.03(5)\\
4.16 &6.369(3) &292.9(1.3)&63.2(1.0) &33.52(15) &5.95(4)\\
7.63 &6.186(3) &278.7(1.7)&61.3(1.3) &34.37(21) &5.80(5)\\
9.92 &6.100(2) &272.2(1.6)&60.7(1.6) &34.77(20) &5.73(4)\\
9.18 &6.087(3) &268.5(1.8)&59.6(1.2) &35.01(21) &5.70(4)\\

 \hline \hline \\

\end{tabular}
   \end{center}
\end{table}

One should note that the observation of a zero pressure effect on
$\lambda(0)$, suggests that almost the whole PE on $\kappa(0)$ in
YB$_6$ is determined by the pressure dependence of $\xi(0)$. The PE
on $\xi(0)$ is twofold. First of all, by neglecting the first term
at the right hand side of Eq.~(\ref{eq:delta-xi}), one can easily
see that the absolute value of the PE on $\xi(0)$ is expected to be
the same as the one on $T_c$. Second, the absolute value of $\xi(0)$
was determined from the measured $H_{c2}(0)$ by means of
Eq.~(\ref{eq:xi}) implying, that the PE on $H_{c2}(0)$ is twice as
big as the PE on $\xi(0)$ and, as a consequence, twice as big as the
PE on $T_c$. The value of $H_{c2}(0)$, however, is not uniquely
determined by $T_c$. Various superconductors, having similar
$T_c$-s, can have the upper critical fields values that are
different by a few orders of magnitude. According to the WHH theory
$H_{c2}(0)\propto T_c \cdot dH_{c2}/dT|_{T=T_c}$ so that the
following relation holds
\begin{equation}
\frac{d\ln H_{c2}(0)}{dp}=\frac{d\ln
[dH_{c2}/dT|_{T=T_c}]}{dp}+\frac{d\ln T_c}{dp} =-2\frac{d\ln
\xi(0)}{dp}.
 \label{eq:Hc2}
\end{equation}
This implies that in superconductors for which the absolute value of
the pressure shift of $T_c$ is much bigger than $1/B$ the pressure
shifts of the superconducting quantities such as $T_c$, $\kappa(0)$,
$\xi(0)$, $H_{c2}(0)$, $dH_{c2}/dT|_{T=T_c}$ are not independent but
related to each other in accordance with:
\begin{eqnarray}
 \nonumber
 \frac{d\ln T_c}{dp}\simeq
 \frac{d\ln\kappa(0)}{dp}\simeq
 -\frac{d\ln\xi(0)}{dp}\simeq
 0.5 \frac{d\ln H_{c2}(0)}{dp}\simeq\\
 \simeq\frac{d\ln(dH_{c2}/dT|_{T=T_c})}{dp}. \ \ \ \
 \label{eq:relations}
\end{eqnarray}
The YB$_6$ superconductor [$1/B=1/1900$kbar =0.05\%/kbar
\cite{Zirngiebl86}, $d\ln T_c/dp=$-0.83(1)\%/kbar (see inset in
Fig.~\ref{fig:TcHc2Lam}(a) and Table~\ref{Table:pressure_results})]
satisfy the above mentioned criteria. This implies that for YB$_6$
the relation~(\ref{eq:relations}) is expected to be correct. A quick
glance at  Fig.~\ref{fig:relative_shifts}(b) and
Table~\ref{Table:pressure_results} reveals that this is exactly the
case. All the quantities entering Eq.~(\ref{eq:relations}) are
almost equal to each other within the accuracy of the experiment.

A good agreement of the experimental data with the simple above
presented approach suggests that in the BCS superconductors, where
condition $d\ln T_c/dp \gg 1/B$ holds, one needs to measure the
pressure effect on $T_c$ only. The pressure dependences of
$H_{c2}(0)$, $dH_{c2}/dT|_{T=T_c}$, $\xi(0)$ and $\kappa(0)$ can be
obtained then by using Eq.~(\ref{eq:relations}). In order to check
the validity of Eq.~(\ref{eq:relations}), similar pressure
experiments were performed for RbOs$_2$O$_6$ \cite{Khasanov06unp}.
We have also analyzed the $H_{c2}$ vs. $p$ dependence for MgB$_2$
from Ref.~\cite{Suderow04}. In both cases a good agreement between
the experimental data and Eq.~(\ref{eq:relations}) was observed.

To summarize, measurements of the pressure effect on the transition
temperature $T_c$, the second critical field $H_{c2}$ and the
magnetic penetration depth $\lambda$ were performed on the YB$_6$
superconductor. It was obtained that $T_c$, $H_{c2}(0)$ and the
coherence length $\xi(0)\propto H_{c2}(0)^{-1/2}$ change linearly
with pressure. The pressure coefficients for the superconducting
transition temperature $T_c$, for the second critical field
$H_{c2}(0)$ and for the coherence length $\xi(0)$ turned out to be
$dT_c/dp=-0.0547(4)$~K/kbar, $\mu_0dH_{c2}(0)/dp=-4.84(20)$~mT/kbar
and $d\xi(0)/dp=0.28(2)$~nm/kbar, respectively. No pressure effect
on $\lambda(0)=199.4(1.5)$ was observed within the experimental
accuracy. This implies that the Ginzburg-Landau parameter
$\kappa(0)=\lambda(0)/\xi(0)$ is pressure dependent, decreasing from
6.17(5) at ambient pressure to 5.70(4) at $p=9.2$~kbar. Thus,
pressure {\it softens} the YB$_6$ superconductor and drives it in
the type-I direction. It was also shown that in BCS superconductors
for which the absolute value of the pressure shift of $T_c$ is much
bigger than $1/B$, pressure induced shifts of the superconducting
quantities such as $T_c$, $\kappa(0)$, $\xi(0)$, $H_{c2}(0)$,
$dH_{c2}/dT|_{T=T_c}$ are related to each other.

The authors are grateful to I.L.~Landau for useful comments and
dicussions. This work was supported by the Swiss National Science
Foundation.


\begin{thebibliography}{99}
%
\bibitem{Tinkham75} M.~Tinkham, ''Introduction to
Superconductivity``, {\it Krieger Publishing company, Malabar,
Florida, 1975}.
%
\bibitem{Chandrasekhar93} B.S.~Chandrasekhar and D.~Einzel,
Annalen der Physik {\bf 2}, 535 (1993).
%
\bibitem{Hirschfeld93} P.~J. Hirschfeld and N.~Goldenfeld,
Phys.~Rev.~B {\bf 48}, 4219 (1993).
%
\bibitem{Eiling81} A.~Eiling and J.S.~Schilling, J.~Phys.~F {\bf 11}, 623 (1981).
%
\bibitem{webelements} {\it www.webelements.com}.
%
%
\bibitem{Lortz06} R.~Lortz, Y.~Wang, U.~Tutsch, S.~Abe, C.~Meingast,
P.~Popovich, W.~Knafo, N.~Shitsevalova, Yu.B.~Paderno, and A.~Junod,
Phys.~Rev.~B {\bf 73}, 024512 (2006).
%
\bibitem{Straessle02} T.~Straessle, Ph.D thesis, ETH Zurich, 2001.
%
\bibitem{Werthamer66} E.~Helfand and N.R.~Werthamer,
Phys.~Rev. {\bf 147}, 288 (1966);
N.R.~Werthamer, E.~Helfand, and P.C.~Hohenberg, {\it ibid} {\bf 147}, 295 (1966).
%
\bibitem{Shoenberg40} D.~Shoenberg, Proc.~R.~Soc.~Lond. {\bf A 175},
49 (1940).
%
\bibitem{Khasanov06unp} R.~Khasanov {\it et al.} unpublished.
%
\bibitem{Schneider87} R.~Schneider, J.~Geerk, H.~Reitschel,
Europhys.~Lett. {\bf 4}, 845 (1987).
%
\bibitem{Zirngiebl86} E.~Zirngiebl, S.~Blumenr\"oder, R.~Mock, and G.~G\"untherodt,
J.~Magn.~Magn.~Mater. {\bf 54–57}, 359 (1986).
%
\bibitem{Khasanov04} R.~Khasanov, D.G.~Eshchenko, J.~Karpinski,
S.M.~Kazakov, N.D.~Zhigadlo, R.~Br\"utsch, D.~Gavillet,
D.~Di~Castro, A.~Shengelaya, F.~La~Mattina, A.~Maisuradze,
C.~Baines, and H.~Keller, Phys.~Rev.~Lett. {\bf 93}, 157004 (2004).
%
\bibitem{Khasanov05} R.~Khasanov, D.Di~Castro, M.~Belogolovskii,
Yu.~Paderno, V.~Filippov, R.~Br\"utsch, and H.~Keller, Phys.~Rev.~B
{\bf 72}, 224509 (2005).
%
\bibitem{Suderow04} H.~Suderow, V.G.~Tissen, J.P.~Brison, J.L.~Martinez,
S.~Vieira, P.~Lejay, S.~Lee, and S.~Tajima, Phys.~Rev.~B {\bf 70},
134518 (2004).
%
\end{thebibliography}
\end{document}